\documentclass[%
 aip,
 jcp,%
 amsmath,amssymb,
preprint,%
]{revtex4-1}

\usepackage{graphicx}
\usepackage{dcolumn}
\usepackage{bm}
\usepackage{color}

\begin{document}

\title{Implementation of Lees-Edwards periodic boundary conditions for
direct numerical simulations of
particle dispersions under shear flow}

\author{Hideki Kobayashi}
\email{hidekb@cheme.kyoto-u.ac.jp}
\author{Ryoichi Yamamoto}
\email{ryoichi@cheme.kyoto-u.ac.jp}
\affiliation{Department of Chemical Engineering, Kyoto University, Kyoto
615-8510, Japan}
\affiliation{CREST Japan Science and Technology Agency,
Kawaguchi 332-0012, Japan}

\date{\today}

\begin{abstract}
 A general methodology is presented to perform direct numerical simulations
 of particle dispersions in a shear flow with Lees-Edwards periodic boundary conditions.
 The Navier-Stokes equation is solved in oblique coordinates to resolve the incompatibility of the
 fluid motions with the sheared geometry, and the force
 coupling between colloidal particles and the host fluid is imposed by 
 using a smoothed profile method.
 The validity of the method is carefully examined by comparing the
 present numerical results with experimental viscosity data for
 particle dispersions in a wide range of volume fractions and shear rates
 including nonlinear shear-thinning regimes.
\end{abstract}

\pacs{83.80.Rs, 47.57.Ng, 83.80.Hj, 83.10.Rs}

\keywords{colloidal dispersion, simulation, hydrodynamic interaction,
shear flow}
\maketitle

\section{introduction}

Understanding the rheological properties of particle dispersions
has been an important problem in many fields of science and engineering. 
When a dispersion is subjected to shear flow, the flow
properties of the dispersion show a variety of non-Newtonian
behaviors such as shear thinning and shear thickening.
These non-Newtonian behaviors are associated with the
changing microstructures of the dispersion, and 
several different physical mechanisms for these peculiar behaviors 
have been proposed.

In recent years, several numerical methods have been developed
to accurately simulate particle dispersions, and they are all based on a similar 
approach, which involves resolving the fluid motion simultaneously
with the particle motion.
We refer to this approach as direct numerical simulation (DNS).
Recently, we have developed a numerical method, which we call 
the smoothed profile method (SPM), for the DNS of particulate flows.
\cite{spm, spm2, spm_fl1, spm_fl2}
In the SPM, the Navier-Stokes equation for the fluid motion is discretized on
a fixed grid, and the Newton's and Euler's equations for the particle
motion are solved simultaneously with the fluid motion. 
One simple technique to impose shear flow with the DNS approach that maintains conventional cubic periodic
boundary conditions is 
to apply a spatially periodic external force to generate a periodic flow profile. 
We have successfully used a zigzag flow profile to impose both steady
and oscillatory shear flows in the DNS of spherical particle dispersions.\cite{spm_fl1,spm_fl2}

When a zero-wavevector shear flow is required, 
the usual cubic periodic boundary conditions must be modified to
be compatible with a time-dependent shear deformation of the simulation
cell.
Such a modification was proposed by Lees and Edwards
\cite{lees-edwards} and is commonly used in various simulation studies.
The Lees-Edwards boundary conditions can be very easily implemented for
particle-based simulations such as molecular dynamics simulations.
However, care must be taken to implement these conditions in continuum grid-based
simulations such as computational fluid dynamics or time-dependent
Ginzburg-Landau equations.
The most useful implementation of the Lees-Edwards periodic boundary
conditions for grid-based simulations is to solve the dynamic equations
in deformed (oblique) coordinates.\cite{onuki, toh_2d_GL, rogallo}
Onuki proposed a general methodology to examine the phase transition
dynamics and rheology in the presence of shear flow,\cite{onuki} 
and it has been successfully used in several simulation studies and particularly for
polymeric fluids in shear flow.
\cite{AYT_2d_GL_poly,zhenli_2d_GL_binary,imaeda_2004,nishitsuji_2010}

The aim of this short paper is to propose a method to implement the
Lees-Edwards periodic boundary conditions to simulate dispersions of
solid particles in host fluids by the combinatory use of the
SPM and the oblique coordinates.

\section{method}

In the SPM, the boundary between the solid particles and the solvent is
 replaced with a continuous interface by assuming a smoothed
 profile. 
This simple modification enables us to calculate hydrodynamic interactions both
 efficiently and accurately without neglecting many-body interactions.
The equation governing the dynamics of particle dispersion
 is a modified Navier-Stokes equation:
 \begin{eqnarray}
  \rho \left\{ \frac{\partial \bm{u}}{\partial t} +
		 (\bm{u}\cdot\bm{\nabla})\bm{u} \right\}
  = \bm{\nabla} \cdot \bm{\sigma} 
  + \rho \phi\bm{f}_{\rm p} 
  - K \rho(u_x - \dot{\gamma}y)\bm{e}_x
  \label{modifiednseq}
 \end{eqnarray}
with the condition of incompressibility $\bm{\nabla}\cdot\bm{u}=0$, where
$\rho$ is the solvent density,
 \begin{eqnarray}
  \bm{\sigma} = -p \bm{I} + \eta_{\rm{f}} \left\{ \bm{\nabla} \bm{u} +
					   (\bm{\nabla}
					   \bm{u})^{T}\right\}
\label{sigma}
 \end{eqnarray}
is the Newtonian stress tensor with a solvent viscosity of $\eta_{\rm{f}}$, and
$\bm{u}(\bm{r},t)$ and $p(\bm{r},t)$
are the velocity and pressure of the dispersion, respectively. 
A smoothed profile function $0\leq\phi(\bm{r},t)\leq1$
 distinguishes between the fluid and particle domains as well as yields $\phi=1$
 in the particle domain and $\phi=0$ in the fluid domain. These domains
 are separated by thin interstitial regions with thicknesses
 characterized by $\xi$. The dispersion density $\rho$ is represented as 
 \begin{eqnarray}
  \rho = (1-\phi)\rho_{\rm f} + \phi \rho_{\rm p}
 \end{eqnarray}
 where $\rho_{\rm f}$ and $\rho_{\rm p}$ are the solvent and particle densities, respectively.
 Only neutral buoyancy dispersions 
 with $\rho=\rho_{\rm f}=\rho_{\rm p}$ are simulated in the present study.
 The body force $\phi\bm{f}_{\rm p}$ is
 introduced so that the total velocity field $\bm{u}$ of the dispersion
 satisfies
$\bm{u}(\bm{r})=(1-\phi)\bm{u}_{f}(\bm{r}) + \phi\bm{u}_{p}(\bm{r})$, where
$\bm{u}_{f}$ is the fluid velocity and $\bm{u}_{p}$
represents the rigid motions of the particles.
The incompressible condition $\bm{\nabla} \cdot \bm{u}$ thus ensures
$\bm{\nabla}\phi \cdot (\bm{u}_{p} - \bm{u}_{f})$
because both $\bm{u}_{f}$ and $\bm{u}_{p}$ satisfy incompressible conditions.
The gradient of $\phi$ is proportional to the surface-normal vector
and has a support on the interfacial domains. Therefore, 
the body force $\phi\bm{f}_{\rm p}$ introduced to satisfy
the rigidity of the particles ensure the appropriate 
impermeability boundary conditions at the fluid-particle interface,
while the non-slip boundary conditions are imposed automatically by the viscous stress
 term in the Navier-Stokes equation.
More detailed explanations and
 the mathematical expressions for $\phi$ and $\phi\bm{f}_{\rm p}$ were
 also detailed in our previous papers.\cite{spm,spm2}

 The last term in Eq.(\ref{modifiednseq}) represents the external force needed 
to maintain linear shear flow:
 \begin{eqnarray}
  u_x=
   \dot{\gamma}y
\label{linear}
 \end{eqnarray}
 where $\dot{\gamma}$ is the shear rate, and $K$ is a constant that
 determines the amplitude of the external force.
Here we impose only the zero-wavevector shear flow
so that the averaged fluid velocity becomes compatible with Eq.(\ref{linear}).

The motion of the $i$-th particle in a dispersion is governed by Newton's
 and Euler's equations of motion:
 \begin{eqnarray}
  M_i\frac{d}{dt}\bm{v}_i=\bm{f}_i^{\rm H}+\bm{f}_i^{\rm P}+\bm{g}_i^V,\;\;\;
   \frac{d}{dt}\bm{r}_i=\bm{v}_i
   \label{newtoneq}
 \end{eqnarray}
 \begin{eqnarray}
  \bm{I}_i\cdot\frac{d}{dt}\bm{\omega}_i=\bm{n}_i^{\rm
   H}+\bm{g}_i^{\omega}
   \label{eulereq}
 \end{eqnarray}
 where $\bm{r}_i$, $\bm{v}_i$, and $\bm{\omega}_i$ are the position,
 translational velocity, and rotational velocity of the colloidal particles,
 respectively. $M_i$ and $\bm{I}_i$ are the mass and the moment of inertia,
 and $\bm{f}_i^{\rm H}$ and $\bm{n}_i^{\rm H}$ are the hydrodynamic
 force and torque exerted by the solvent on the colloidal particles, 
respectively
 \cite{spm,spm2}. $\bm{g}_i^{\rm v}$ and $\bm{g}_i^{\omega}$ are the
 random force and torque, respectively, due to thermal fluctuations. The
 temperature of the system is defined such that the long-term diffusive
 motion of the colloidal particles reproduces the Stokes-Einstein rule.
 \cite{spm_fl1,spm_fl2}
$\bm{f}_i^{\rm P}$ represents the potential force due to direct
 inter-particle interactions such as through the Coulombic and Lennard-Jones
 potentials.

\begin{figure}[t]
  \begin{center}
   \includegraphics[width=0.8\hsize]{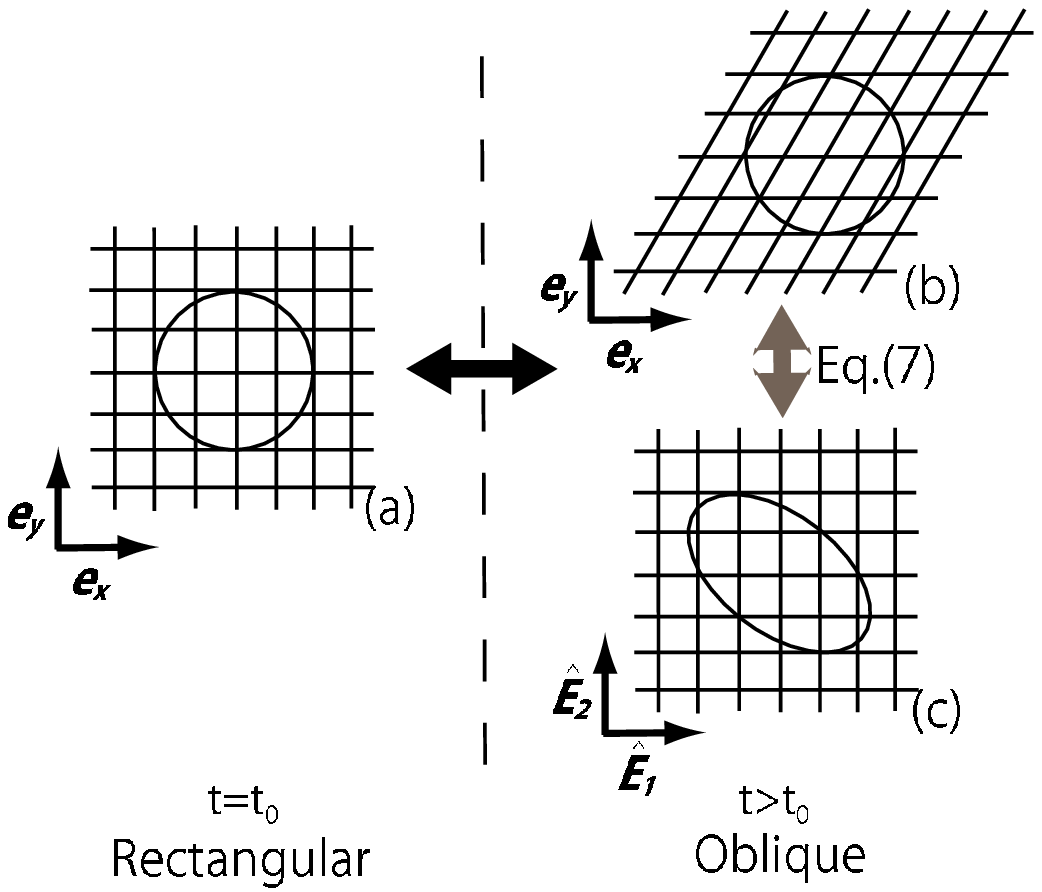}
   \caption{
A schematic illustration of the present coordinate transformation. In (a), a
   spherical solid particle is in a solvent, which is discretized into
   grids in an ordinary rectangular coordinate system, at a time $t=t_0$.
Since the shear flow is applied for $t>t_0$, the solvent (grids) is
   convected by the flow while the shape of the solid particle
   is unchanged. Such a situation is depicted in the original
   (experimental) 
frame in (b) and also in a transformed (oblique) frame in (c).
The transformation between (b) and (c) is defined by Eq. (\ref{trans}).
\label{transform}
   }
  \end{center}
 \end{figure}

Eqs.(\ref{modifiednseq}), (\ref{newtoneq}), 
and (\ref{eulereq}) are solved simultaneously in the SPM. However,  
this task is not easy with an ordinary periodic boundary
condition because Eq.(\ref{modifiednseq}) depends explicitly on $y$, which 
leads to a violation of the translational invariance. 
This problem can be eliminated by using oblique coordinates.
Fig. \ref{transform} represents a schematic illustration of the present 
coordinate transform. At a time $t=t_0$, a spherical solid particle is
located in a solvent in Fig. \ref{transform} (a) where the solvent is 
discretized into square grids in an ordinary rectangular coordinate system.
In Fig. \ref{transform} (b), the grids are deformed due to the shear flow that is
applied for $t>t_0$ while the shape of the solid particle is unchanged. 
The same situation is depicted in a transformed (oblique) frame in
Fig. \ref{transform} (c) where the grid has not moved (i.e., it remains rectangular), 
but the shape of the solid particle changes over time due to the shear flow.

\begin{figure}[t]
  \begin{center}
   \includegraphics[width=0.7\hsize]{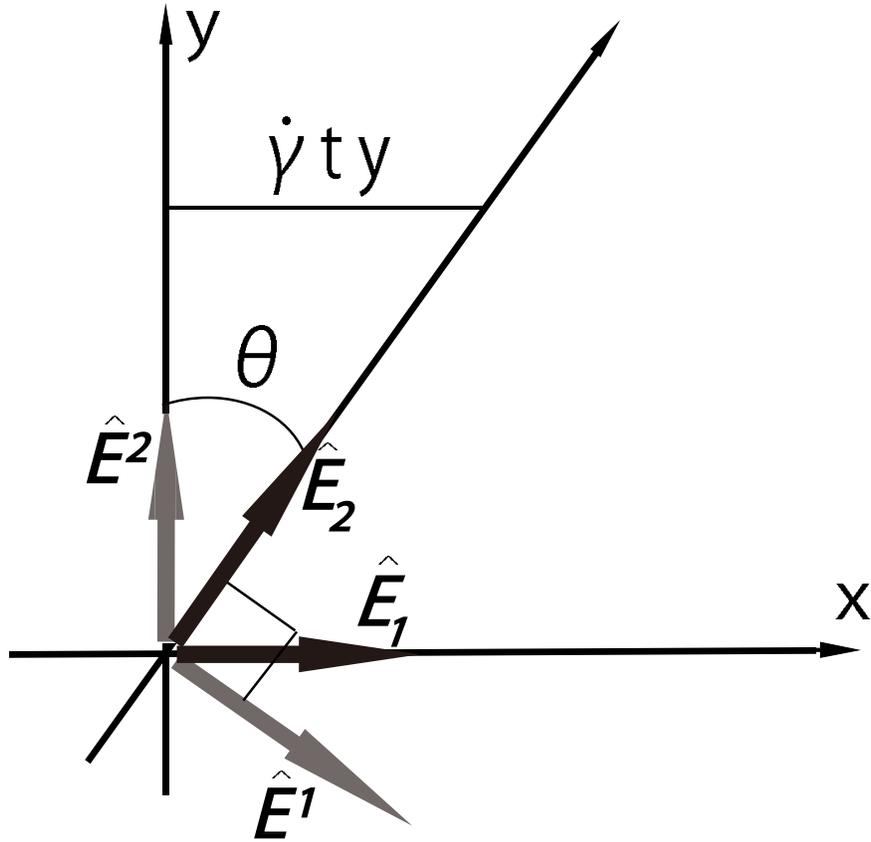}
   \caption{
   The definition of a basis vector.
   \label{coordinate}
   }
  \end{center}
 \end{figure}

To formulate the oblique coordinate transformation based on tensor
 analysis, 
we began by redefining the covariant basis $\hat{\bm{E}}_{i}$ and contravariant
 basis $\hat{\bm{E}}^{i}$ in oblique coordinates rather than using the
 expressions shown in the literature.
\cite{onuki, toh_2d_GL, rogallo}
Fig. \ref{coordinate}
 provides a definition of the basis vectors. Using a rectangular unit vector,
 $\hat{\bm{E}}_{i}$ and $\hat{\bm{E}}^{i}$ are expressed as
 \begin{eqnarray}
  \begin{array}{ll}
   \hat{\bm{E}}_{1}=\bm{e}_{x}
    & \hat{\bm{E}}^{1}=\bm{e}_{x} - \dot{\gamma} t \bm{e}_{y}\\
   \hat{\bm{E}}_{2}=\dot{\gamma} t \bm{e}_{x} + \bm{e}_{y}
    & \hat{\bm{E}}^{2}=\bm{e}_{y}\\
   \hat{\bm{E}}_{3}=\bm{e}_{z} & \hat{\bm{E}}^{3}=\bm{e}_{z}
  \end{array}
\label{trans}
 \end{eqnarray}
 where $\bm{e}_{\alpha}$ is the unit vector in the $\alpha$(= $x,y,z$) direction 
 in the original rectangular coordinate system. 
We can obtain contravariant (covariant) vector components $A^{i}$ 
 ($A_{i}$) using $\bm{A} \cdot \hat{\bm{E}}^{i}$ ($\bm{A} \cdot \hat{\bm{E}}_{i}$).
 The positional vector
 $\bm{r}\equiv x \bm{e}_{x} + y \bm{e}_{y} + z \bm{e}_{z}$ is
 transformed from the rectangular coordinate expression $\bm{r}$ to the oblique
 coordinate expression $\hat{\bm{r}}$ as follows:
 \begin{eqnarray}
  \begin{split}
   \bm{r} &\equiv x \bm{e}_{x} + y \bm{e}_{y} + z \bm{e}_{z}\\
   &=
   (\bm{r}\cdot\hat{\bm{E}}^{1})\hat{\bm{E}}_{1}+
   (\bm{r}\cdot\hat{\bm{E}}^{2})\hat{\bm{E}}_{2}+
   (\bm{r}\cdot\hat{\bm{E}}^{3})\hat{\bm{E}}_{3}\\
   &=
   \hat{x}^1\hat{\bm{E}}_{1}+\hat{x}^2\hat{\bm{E}}_{2}+\hat{x}^3\hat{\bm{E}}_{3}
   \equiv \hat{\bm{r}},
  \end{split}
 \end{eqnarray}
 where the contravariant components ($\hat{x}^{1},\hat{x}^{2},\hat{x}^{3}$)
 are expressed as
 \begin{eqnarray}
  \begin{array}{l}
   \hat{x}^{1}=x - \dot{\gamma} t y\\
   \hat{x}^{2}=y\\
   \hat{x}^{3}=z
   \label{transformationRtO}
  \end{array}
 \end{eqnarray}
 and the time in oblique coordinates is expressed as $\hat{t} = t$. 
Each contravariant
 component is transformed to a covariant component by 
 using the metric tensors  $G_{ij}=\hat{\bm{E}}_i \cdot \hat{\bm{E}}_j$
and $G^{ij}=\hat{\bm{E}}^i \cdot \hat{\bm{E}}^j$. 
Then, the transformation can be expressed as
 \begin{eqnarray}
  A^{i} = G^{ij}A_{j}\\
  A_{i} = G_{ij}A^{j}
 \end{eqnarray}
 
The physical quantities in Eq. (\ref{modifiednseq}) are transformed as
 indicated below:
 \begin{eqnarray}
  \hat{p}(\hat{\bm{r}},\hat{t}) = p(\bm{r},t)
 \end{eqnarray}
 \begin{eqnarray}
   \hat{\phi}(\hat{\bm{r}},\hat{t}) = \phi(\bm{r},t)
 \end{eqnarray}
 \begin{eqnarray}
  \hat{\bm{u}}(\hat{\bm{r}},\hat{t}) = \bm{u}(\bm{r},t) - \dot{\gamma}y \bm{e}_{x}
 \end{eqnarray}
 \begin{eqnarray}
  \hat{\phi}\hat{\bm{f}}_{\rm{p}}(\hat{\bm{r}},\hat{t}) = \phi\bm{f}_{\rm{p}}(\bm{r},t)
   \label{transformfp}
 \end{eqnarray}
In the oblique coordinate system, $\hat{\bm{u}}$ satisfies 
the standard periodic boundary conditions while $\bm{u}$ satisfies the Lees-Edwards
 periodic boundary conditions in the rectangular coordinate system.
 The contravariant components ($\hat{u}^{1},\hat{u}^{2},\hat{u}^{3}$) of
 $\hat{\bm{u}}$ are expressed as
 \begin{eqnarray}
  \begin{array}{l}
   \hat{u}^{1}= u_x - \dot{\gamma} t u_y - \dot{\gamma} y\\
   \hat{u}^{2}= u_y\\
   \hat{u}^{3}= u_z
   \label{transformationRtOv}
  \end{array}
 \end{eqnarray}
 where ($u_x,u_y,u_z$) are the rectangular components of $\bm{u}$. 
The contravariant components of $\hat{\phi}\hat{\bm{f}_{\rm{p}}}$ are 
 ($\hat{\phi}\hat{f}_{\rm{p}}^{1},\hat{\phi}\hat{f}_{\rm{p}}^{2},\hat{\phi}\hat{f}_{\rm{p}}^{3}$) and can be
expressed as
 \begin{eqnarray}
  \begin{array}{l}
   \hat{\phi}\hat{f_{\rm{p}}}^{1}= \phi f_{\rm{p}}^x - \dot{\gamma} t
    \phi f_{\rm{p}}^y\\
   \hat{\phi}\hat{f_{\rm{p}}}^{2}= \phi f_{\rm{p}}^y\\
   \hat{\phi}\hat{f_{\rm{p}}}^{3}= \phi f_{\rm{p}}^z
   \label{transformationRtOfp}
  \end{array}
 \end{eqnarray}
 where ($f_{\rm{p}}^x,f_{\rm{p}}^y,f_{\rm{p}}^z$) are the rectangular
 components of $\bm{f}_{\rm{p}}$.

The differential operators in oblique coordinates are defined by
 \begin{eqnarray}
  \hat{\bm{\nabla}} = \hat{\bm{E}}^{1} \frac{\partial}{\partial \hat{x}^{1}} +
   \hat{\bm{E}}^{2} \frac{\partial}{\partial \hat{x}^{2}} +
   \hat{\bm{E}}^{3} \frac{\partial}{\partial \hat{x}^{3}}
\label{nabla} 
\end{eqnarray}
 \begin{eqnarray}
  \frac{\partial}{\partial \hat{t}}= \frac{\partial}{\partial t} + 
   \dot{\gamma} y \frac{\partial}{\partial x}
 \end{eqnarray}
Therefore, the Laplacian operator in oblique coordinates is expressed as
 \begin{eqnarray}
  \begin{split}
   \hat{\bm{\Delta}} &= \left(\frac{\partial}{\partial \hat{x}^1} \right)^2 +
   \left(\frac{\partial}{\partial \hat{x}^2} - \dot{\gamma} t
   \frac{\partial}{\partial \hat{x}^1}  \right)^2 +
   \left(\frac{\partial}{\partial \hat{x}^3} \right)^2
  \end{split}
  \label{laplacian}
 \end{eqnarray}

Using these formula, Eqs. (\ref{modifiednseq}) and (\ref{sigma}) are
rewritten in oblique coordinates as
 \begin{eqnarray}
  \begin{split}
   \rho  \left\{ \frac{\partial \hat{\bm{u}}}{\partial \hat{t}}+ 
   (\hat{\bm{u}}\cdot\hat{\bm{\nabla}})\hat{\bm{u}} \right\}&\\
   = \hat{\bm{\nabla}} \cdot \hat{\bm{\sigma}} 
   + \rho \hat{\phi}\hat{\bm{f}_{\rm p}} 
   - &\rho \dot{\gamma} \hat{u}^2 \hat{\bm{E}}_{1} 
   - K \rho (\hat{u}^{1} + \gamma \hat{u}^{2})\hat{\bm{E_1}}
  \end{split}
  \label{obliquenseq}
 \end{eqnarray}
and 
 \begin{eqnarray}
  \begin{split}
  \hat{\sigma}^{ij}(\hat{\bm{r}},\hat{t})=
   -G^{ij}\hat{p}(\hat{\bm{r}},\hat{t})& +
   \eta_{\rm{f}} \left\{
   G^{in}\frac{\partial \hat{u}^j}{\partial \hat{x}^n} +
   G^{jm}\frac{\partial \hat{u}^i}{\partial \hat{x}^m}
   \right\}
  \end{split}
  \label{oblique_stress_tensor}
 \end{eqnarray}
 with the incompressibility condition
 $\hat{\bm{\nabla}}\cdot\hat{\bm{u}}=0$. 
Because Eq. (\ref{obliquenseq}) and $\hat{\bm{u}}$ satisfy the periodic boundary
 conditions in all directions, a fast Fourier
 transformation (FFT) can be safely used to solve the Poisson equation, which is needed to determine
 $\hat{p}$ with the incompressibility condition.
In Appendix 1, detailed explanations are given on how to solve 
Eq. (\ref{obliquenseq}) with the incompressibility condition in
the oblique coordinate system with using the spectral (Fourier)
method.

 When $\gamma\equiv\dot{\gamma}t = 1$,
 the positions
 $\hat{\bm{r}}=(\hat{x}^1,\hat{x}^2,\hat{x}^3)$
 on an oblique grid with $\gamma$
 can be mapped onto the identical positions
 $\bm{r}=(x,y,z)$ on the original rectangular grid with $\gamma=0$
 using the operation 
$x=\hat{x}^1+\hat{x}^2$, $y=\hat{x}^2$, $z=\hat{x}^3$.
 The shear strain $\gamma$ is then reset to 0.\cite{onuki}
 Repeating this process allows us to perform stable numerical calculations
 over a long period with keeping $0\le\gamma\le1$.
 The above coordinate transformation based on the tensor analysis leads to
 the same expression for the Laplacian $\hat{\bm{\Delta}}$ as that of a
 previous study.\cite{onuki}
 However, a difference arises between the differential 
 operators $\hat{\nabla}$ for which our formal transformation derives a
 much simpler expression as shown in Eq. (\ref{nabla}).

We calculate the dynamics of solid particle dispersions in shear
flow by following these steps: 

i) The fluid velocity field in the oblique coordinate system at a new time
$t=nh$ is calculated by integrating Eq. (\ref{obliquenseq}) over time with
$\hat{\phi}\hat{\bm{f}_{\rm p}}=0$ as
 \begin{eqnarray}
  \begin{split}
   &\hat{\bm{u}}^{*}= \hat{\bm{u}}^{n-1} + \\ 
   & \int^{t_{n-1}+h}_{t_{n-1}}\left[
   \hat{\bm{\nabla}} \cdot \left(
   \frac{1}{\rho}\hat{\bm{\sigma}} - \hat{\bm{u}}\hat{\bm{u}}
   \right) -
   \left\{ K (\hat{u}^{1} + \gamma \hat{u}^{2})
   +2\dot{\gamma}\hat{u}^{2}\right\}\hat{\bm{E}}_{1}
   \right] ds
  \end{split}
  \label{time_development_nseq}
 \end{eqnarray}
 while satisfying the incompressibility condition 
$\hat{\bm{\nabla}}\cdot\hat{\bm{u}}^{*} = 0$.
Here, the superscript $n$ denotes the time step, and $h$ is the time
increment. 
In Appendix 1, detailed explanations are given also on how to solve 
Eq. (\ref{time_development_nseq}) with using the spectral method.

ii) The velocity field $\hat{\bm{u}}^{*}$ is transformed into rectangular
 coordinates $\bm{u}^{*}$ using the inverse transformation expressed as
 \begin{eqnarray}
  \begin{array}{l}
   u_x= \hat{u}^{1} + \dot{\gamma} t \hat{u}^{2} + \dot{\gamma} \hat{x}^{2}\\
   u_y= \hat{u}^{2}\\
   u_z= \hat{u}^{3}.
  \end{array}
  \label{velocity_fo_tr}
 \end{eqnarray}

iii) The motions of colloidal particles are only calculated in rectangular
coordinates. 
The position of each colloidal particle is calculated by
 \begin{eqnarray}
   \begin{split}
    \bm{r}^{n}_{i}= \bm{r}^{n-1}_{i} +
     \int^{t_{n-1}+h}_{t_{n-1}} \bm{v}^{n-1}_{i} ds
   \end{split}
 \end{eqnarray}
iv) Using the momentum conservation between colloidal particles and the
solvent, the hydrodynamic force and torque acting on each colloidal
particle are computed with volume integrals within the particle
domain as
 \begin{eqnarray}
  \bm{f}_{i}^H = \frac{\rho}{h} \int \bm{dr} [\phi^{n}_{i} 
   \left(\bm{u}^{*} - \bm{u}^{n-1}_{p} \right)] 
\label{fff}
 \end{eqnarray}
and
 \begin{eqnarray}
  \bm{n}_{i}^H = \frac{\rho}{h} \int \bm{dr} [(\bm{r} -
   \bm{r}_{i})\times\phi^{n}_{i}  
   \left(\bm{u}^{*} - \bm{u}^{n-1}_{p} \right)]
\label{nnn}
 \end{eqnarray}
 where $\phi \bm{u}^{n-1}_{\rm p}(\bm{r}) = \sum_{i} \phi^{n}_{i}(\bm{r}) \left(\bm{v}^{n-1}_{i} + \bm{\omega}^{n-1}_{i} \times\ (\bm{r} - \bm{r}_i) \right)$
 is the correct velocity field within the particle domain in which 
$\phi \simeq 1$.
The space integrals in Eqs.(\ref{fff}) and (\ref{nnn}) are carried out 
by summations over grid points in actual computations, however, there
 occur grid mismatch between $\bm{u}^*$ which is supported on oblique
 grid points $\hat{\bm{r}}_{\hat{i},\hat{j},\hat{k}}$ and 
other variables ($\phi_i^n$ and $\phi\bm{u}_{\rm p}^{n-1}$)
which are supported on rectangular grid points $\bm{r}_{i,j,k}$.
We determine values of $\bm{u}^*$ on rectangular grid points
 $\bm{r}_{i,j,k}$ by
 linear interpolation as described in detail in Appendix 2.
 The translational velocity and rotational velocity of each colloidal
 particle are then calculated as
 \begin{eqnarray}
  \bm{v}^{n}_{i}= \bm{v}^{n-1}_{i} +
   \frac{1}{M_{i}}\int^{t_{n-1}+h}_{t_{n-1}} 
   \left(
    \bm{f}_{i}^H + \bm{f}_{i}^P + \bm{g}_{i}^V
   \right) ds
   \label{HI_translation}
 \end{eqnarray}
and
 \begin{eqnarray}
  \bm{\omega}^{n}_{i}= \bm{\omega}^{n-1}_{i} +
   \bm{I}_{i}^{-1} \int^{t_{n-1}+h}_{t_{n-1}} 
   \left(
    \bm{n}_{i}^H + \bm{g}_{i}^{\omega}
   \right) ds
   \label{HI_torque}
 \end{eqnarray}

v) To ensure the rigidity of the
 particles and the appropriate non-slip boundary conditions at the
 fluid/particle interface, the body force $\phi\bm{f}_{\rm p}$ is
 calculated as
 \begin{eqnarray}
  \phi\bm{f}_{\rm p} = \frac{\phi   
   \left(\bm{u}^{n}_{\rm p} - \bm{u}^{*} \right)}{h} - \frac{1}{\rho} \bm{\nabla}p_{\rm p}.
 \end{eqnarray}
 The correcting pressure $p_{\rm p}$ is determined to make the resultant total
 velocity incompressible. This leads to the Poisson equation of $p_{\rm p}$:
 \begin{eqnarray}
  \bm{\Delta} p_{\rm p} =
   \rho \frac{\bm{\nabla}\cdot\phi\left(\bm{u}^{n}_{\rm p} - \bm{u}^{*} \right)}{h}.
 \end{eqnarray}
We then transform $\phi\bm{f}_{\rm p}$ into oblique coordinates 
$\hat{\phi}\hat{\bm{f}_{\rm p}}$ using Eqs. (\ref{transformfp}) and (\ref{transformationRtOfp}). 

vi) Finally, we obtain the correct fluid velocity field as:
 \begin{eqnarray}
  \hat{\bm{u}}^{n} = \hat{\bm{u}}^{*} + \hat{\phi}\hat{\bm{f}_{\rm p}}h.
   \label{correct_fluid_vector}
 \end{eqnarray}
Repetition of steps i) through vi) provides a complete procedure to perform the DNS of colloidal
dispersions under shear flow.

We can calculate the stress tensor of the dispersion 
 $\langle \bm{s} \rangle$
 and the dispersion viscosity $\eta = \langle s_{xy} \rangle /
 \dot{\gamma}$ in the following manner
 where $\langle \cdots \rangle$ denotes averaging over space and time.
 The equation governing the dispersion is formally
 written as:
 \begin{eqnarray}
   \frac{D}{Dt}(\rho \bm{u})
  = \bm{\nabla} \cdot \bm{\sigma}^{\rm dis} - K \rho(u_x - \dot{\gamma}y)\bm{e}_x
  \label{dispersioneq}
 \end{eqnarray}
 By comparing Eq. (\ref{modifiednseq}) with Eq. (\ref{dispersioneq}), we get
 the formula
 \begin{eqnarray}
 \begin{split}
  \bm{\nabla} \cdot \bm{\sigma}^{\rm dis}=\bm{\nabla} \cdot \bm{\sigma} + \rho \phi \bm{f}_{\rm p}.
  \label{transform_sigma}
  \end{split}
 \end{eqnarray}
 The full stress tensor $\bm{s}$ of the flowing dispersion is then defined
 by introducing a convective momentum-flux tensor explicitly as
 \begin{eqnarray}
  \bm{s} = \bm{\sigma}^{\rm dis} - \rho \bm{u}\bm{u}.
  \label{full_tensor}
 \end{eqnarray}
 The definitions of $\bm{\sigma}^{\rm dis}$ and $\bm{s}$ are
 identical to the definitions in our previous paper.\cite{spm_fl2}
 Now, we can evaluate the average stress tensor of the dispersion 
 $\langle \bm{s} \rangle$ directly from
 Eqs. (\ref{transform_sigma}), (\ref{full_tensor}), and
 $\delta\bm{\sigma}=\bm{s}-\bm{\sigma}$ as
 \begin{eqnarray}
  \begin{split}
   \langle \bm{s} \rangle 
   =& \langle \bm{\sigma} \rangle +
   \frac{1}{V} \left< \int \bm{dr} \delta \bm{\sigma} \right>_{t}\\
   =& \langle \bm{\sigma} \rangle + \frac{1}{V} \left< \int \bm{dr}
   \left[ \left( \bm{\nabla}\cdot(\delta\bm{\sigma}\bm{r}) \right)^{T} -
   \bm{r} \bm{\nabla}\cdot\delta\bm{\sigma} \right] \right>_{t}\\
   =& \langle \bm{\sigma} \rangle 
   -\frac{1}{V} \left< \int \bm{dr} \bm{r}
   \bm{\nabla}\cdot\delta\bm{\sigma} \right>_{t}\\
   =& \langle \bm{\sigma} \rangle 
   -\frac{1}{V} \left< \int \bm{dr} \bm{r} \rho \phi \bm{f}_{\rm p}
   \right>_{t}
   +\frac{1}{V} \left< \int \bm{dr} \bm{r}
   \bm{u}\cdot\bm{\nabla}(\rho \bm{u}) \right>_{t}\\
   =& \langle \bm{\sigma} \rangle 
   -\frac{1}{V} \left< \int \bm{dr} \bm{r} \rho \phi \bm{f}_{\rm p}
   \right>_{t}
  \end{split}
 \end{eqnarray}
 with the volume $V = L_xL_yL_z$ where $L_i$ is the system size in
 $i$-direction.
 $\langle \cdots \rangle_{t}$ denotes time averaging over
 steady state.
 In the derivation of the second formula, we used
 a second rank identity. If we substitute Eq. (\ref{transform_sigma}) into
 the third formula, then we obtain the fourth formula.
 The fifth formula can be obtained by assuming that the system is in
 a steady state in which 
$\left< \frac{d}{dt}\left(\rho \bm{u} \right) \right>_{t} =
   \left< \frac{\partial}{\partial t}\left(\rho \bm{u} \right)
    + \bm{u}\cdot\bm{\nabla}(\rho \bm{u})\right>_{t} =0$ 
and 
  $\left< \frac{\partial}{\partial t}\left(\rho \bm{u} \right)\right>_{t}=0$.

\section{results \label{main_results}}

 Using the method described above, we calculated the high- and low-shear limiting viscosities of colloidal dispersions for various volume
 fractions of particles $\Phi$. The particles
 interact via a truncated
 Mie ($m,n$) potential with $m=36$ and $n=18$. \cite{mie_1903}
 \begin{eqnarray}
  U(r)=\left\{
	\begin{array}{ll}
	 4\epsilon \left\{
		    \left( \dfrac{\sigma}{r}\right)^{36} -
		    \left( \dfrac{\sigma}{r}\right)^{18}\right\}
	 + \epsilon &(r<2^{\frac{1}{18}}\sigma), \\
	 0 &(r>2^{\frac{1}{18}}\sigma),
	\end{array}
       \right.
 \end{eqnarray}
where $r$ is the distance between the centers of a pair of particles. 
The parameter $\epsilon$
 characterizes the strength of the interactions, and $\sigma$ represents
 the diameter of the colloidal particles. 
 The lattice spacing $\delta x$ is taken to
 be the unit of length. The unit of time is given by
 $\rho_{\rm f}\delta x^2/\eta$ where $\eta=1$ and $\rho_{\rm
 f}=\rho_{\rm p}=1$.
 The system size is $L_x\times L_y\times L_z=64\times64\times64$.
 Other parameters are set as follows: $\sigma=8$, $\xi=2$, $\epsilon=1$,
 $\eta=1$, $M_i=\pi \sigma^3/6$, and $h=0.067$. The temperature is
 $k_{\rm B}T=7$.  The range of shear rate is
$1.0\times10^{-4}<\dot{\gamma}<0.1$.
 
\begin{figure}[t]
  \begin{center}
   \includegraphics[width=1.1\hsize]{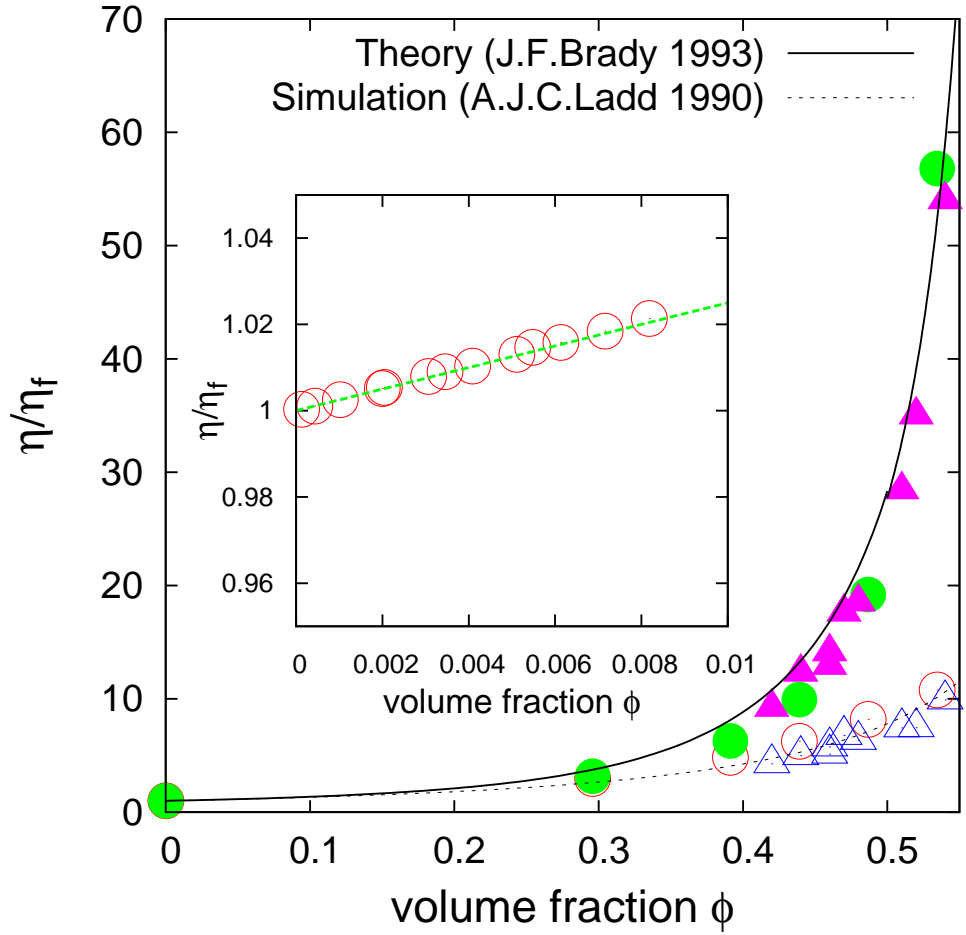}
   \caption{
   The behavior of the viscosity $\eta$ as a function of the volume fraction
   $\Phi$. The open symbols represent the high-shear limiting
   viscosity, and the closed symbols represent the low-shear limiting
   viscosity. The open and closed circles correspond to our 
simulation data, whereas the triangles correspond to experimental 
results.\cite{eta_exp1} 
The solid line is Brady's theoretical 
prediction,\cite{low_eta_theory} and 
the dotted line is a fitting curve obtained from previous
   experimental \cite{eta_exp1} and simulation \cite{high_eta_sim} data.
   The inset indicates a comparison of our present simulation data 
with Einstein's viscosity law (dashed-line) in the small volume fraction regime 
$\Phi<0.01$ where the viscosity exhibits simple Newtonian behavior.
   \label{phi_vs_eta}
   }
  \end{center}
 \end{figure}

 The inset of Fig. \ref{phi_vs_eta} shows 
the dependence of the Newtonian viscosity on the volume 
fraction $\Phi$ when $\Phi\ll1$.
The present simulation data show very good agreement with 
Einstein's viscosity law.
 Fig. \ref{phi_vs_eta} shows the dependence of the low-shear limiting
 viscosity (closed symbols) and 
the high-shear limiting viscosity (open symbols) on the volume
 fraction. 
Our simulation data for both high- and low-shear limiting viscosities
show good agreement with the experimental results of van der Werff 
{\it et al.}\cite{eta_exp1}
 Previously, Brady theoretically predicted the behavior of the low-shear
 limiting viscosity.\cite{low_eta_theory}
Our simulation data show good agreement
 with Brady's prediction over a wide range of volume fractions $0<\Phi<0.55$. 
 Ladd analyzed the behavior of the high-shear limiting viscosity
 using Stokesian dynamics.\cite{high_eta_sim} 
Our simulation data agree well with Ladd's simulation data and also with the
theoretical results of Beenakker.\cite{high_eta_theory}

Finally, we add some comments on the differences between the present 
method using Lees-Edwards boundary condition and the previously
proposed method using zigzag velocity profile.\cite{compare}
We simulated a single spherical particle in shear flow using the two
methods without thermal fluctuation. The volume fraction is $0.001$.
Figure \ref{compare_frq} shows the ratio of angular velocity $\omega$ 
of a spherical particle to the applied shear rate $\dot{\gamma}$ 
as a function of $\dot{\gamma}$. 
Although the data using the present method tend to be slightly smaller
than the data using the zigzag flow, deviations of both data from the analytical 
value $\omega / \dot{\gamma}=0.5$ remain small within numerical errors
of the methods.
Figure \ref{compare_iv} shows the intrinsic viscosity
$[\eta]$ of the dilute dispersion as a function of shear 
rate $\dot{\gamma}$. 
The simulation data using the present method almost perfectly follow onto 
the Einstein's prediction $[\eta]=2.5$, while the data using zigzag
flow slightly overestimate $[\eta]$ because of unphysical kinks of 
the zigzag flow profile.
This problem is not very serious when the shape of dispersed particles
is spherical and the size of the particle is much smaller than the
distance between two kinks.
Serious problems, however, occur if this
method is applied to non-dilute dispersions of chains or rods, for example. 
The present method using  Lees-Edwards boundary condition is free from
this problem.

\section{conclusion}

 We presented a generic methodology for performing DNS of particle
 dispersions in a shear flow using oblique coordinates and 
 periodic boundary conditions.
 The validity of the method was confirmed by comparing the
 present numerical results with experimental viscosity data for
 particle dispersions over a wide range of the parameters $\Phi$ and $\dot{\gamma}$
 that include nonlinear shear-thinning regimes.
 An important advantage of the DNS approach over other approaches such as Stokesian
 dynamics is its applicability to particle dispersions in complex
 fluids.
In fact, electrophoresis of charged colloids \cite{electrophoresis}
 and particle dispersions in nematic liquid crystals
 \cite{colloid_in_nematic} have already been calculated using SPM. 
Our methodology can also be applied to simulate particle dispersions in
viscoelastic fluids simply by replacing the Newtonian constitutive equation
 to more complex ones such as Maxwell model.

\section*{Acknowledgments}

The authors would like to express their gratitude to Dr. T. Murashima,
Dr. Y. Nakayama, Dr. K. Kim, and Dr. T. Iwashita for useful comments and
discussions.

\section*{Appendix 1}

In this section, we describe how to solve Eq. (\ref{obliquenseq}) with
the incompressibility condition in an oblique coordinate system using
Fourier spectral methods. 
The Fourier and inverse Fourier transforms are defined as

\begin{eqnarray}
A(\hat{\bm{k}}) =
 \int A(\hat{\bm{r}})
 \exp (-i \hat{\bm{k}} \cdot \hat{\bm{r}})
 d \hat{\bm{r}}\\
A(\hat{\bm{r}}) =
 \frac{1}{(2 \pi)^3}\int A(\hat{\bm{k}})
 \exp (i \hat{\bm{k}} \cdot \hat{\bm{r}})
 d \hat{\bm{k}},
\end{eqnarray}
where $\hat{\bm{k}}$ is the wavevector of the oblique coordinate system. In $\hat{\bm{k}}$ space, we can express the spatial covariant derivative as
\begin{eqnarray}
\frac{\partial A(\hat{\bm{r}})}{\partial \hat{x^{\alpha}}}
 \rightarrow
 i k_{\alpha}A(\hat{\bm{k}}).
\end{eqnarray}
where $k_{\alpha}$ is a covariant component of $\hat{\bm{k}}$.

Using these relations, we modify Eq. (\ref{time_development_nseq}) from
$\hat{\bm{r}}$ space to $\hat{\bm{k}}$ space. This equation is solved in
$\hat{\bm{k}}$ space. First, by substituting
Eq. (\ref{oblique_stress_tensor}) into
Eq. (\ref{time_development_nseq}), we obtain the explicit equation
represented by 
\begin{eqnarray}
\begin{split}
 \hat{\bm{u}}^{*} (\hat{\bm{r}})=&
 \hat{\bm{u}} (\hat{\bm{r}}) + \\
 \int^{t_{n-1}+h}_{t_{n-1}} & \left[
 - (\hat{\bm{u}}(\hat{\bm{r}}) \cdot \hat{\bm{\nabla}})\hat{\bm{u}}(\hat{\bm{r}})
 - \hat{\bm{\nabla}} \frac{\hat{p}(\hat{\bm{r}})}{\rho}
 + \nu \hat{\bm{\Delta}} \hat{\bm{u}} (\hat{\bm{r}})
 - \left\{ K (\hat{u}^{1}(\hat{\bm{r}}) + \gamma \hat{u}^{2}(\hat{\bm{r}}))
 + 2 \dot{\gamma}\hat{u}^{2}(\hat{\bm{r}})\right\}\hat{\bm{E}}_{1}
 \right] ds,
\end{split}
\label{new_time_development}
\end{eqnarray}
where $\nu=\eta_{\rm f} / \rho$ is the kinetic viscosity.
Using a Fourier transform, the form of Eq. (\ref{new_time_development}) in $\hat{\bm{k}}$ space is written as
\begin{eqnarray}
\begin{split}
 \hat{\bm{u}}^{*}(\hat{\bm{k}})& =
 \hat{\bm{u}} (\hat{\bm{k}}) +\\
 &\int^{t_{n-1}+h}_{t_{n-1}}\left[
 -\bm{F}(\hat{\bm{k}})
 - \nu \hat{\bm{k}}^2 \hat{\bm{u}}(\hat{\bm{k}})
 - \left\{ K (\hat{u}^{1}(\hat{\bm{k}}) + \gamma \hat{u}^{2}(\hat{\bm{k}}))
 + 2 \dot{\gamma}\hat{u}^{2}(\hat{\bm{k}})\right\}\hat{\bm{E}}_{1}
 \right]_{\perp} ds,
\end{split}
\label{k_space_nseq}
\end{eqnarray}
where 
\begin{eqnarray}
\bm{F}(\hat{\bm{k}}) =
 \int (\hat{\bm{u}} (\hat{\bm{r}}) \cdot\hat{\bm{\nabla}})\hat{\bm{u}}
 (\hat{\bm{r}})
 \exp(-i \hat{\bm{k}} \cdot \hat{\bm{r}}) d \hat{\bm{r}}.
 \label{non_liner_term}
\end{eqnarray}
The bracket 
$\left[ \bm{A}(\hat{\bm{k}})\right]_{\perp}
\left(\equiv\bm{A}(\hat{\bm{k}}) \cdot 
 \left(\bm{I} - \frac{\hat{\bm{k}} \hat{\bm{k}}}{\hat{\bm{k}}^2} \right)
\right) $
denotes taking the orthogonal part to $\hat{\bm{k}}$ and this operation corresponds to imposing the incompressibility condition $\hat{\bm{\nabla}} \cdot \hat{\bm{u}}(\hat{\bm{r}})=0$ 
(or equivalently $\hat{\bm{k}} \cdot
\hat{\bm{u}}(\hat{\bm{k}})=0$). Because $\hat{p}(\hat{\bm{r}})$ is
automatically determined by imposing the condition of incompressibility,
we can safely neglect this term.

Using the method describe above, we calculate
$\hat{\bm{u}}^{*}(\hat{\bm{r}})$ from $\hat{\bm{u}}(\hat{\bm{r}})$ as
shown in Eq. (\ref{time_development_nseq}).

\section*{Appendix 2}

%
%

An arbitrary position vector in the oblique coordinate system is defined as
\begin{eqnarray}
\hat{\bm{r}}_{\hat{i},\hat{j},\hat{k}} = (\hat{i} \hat{\bm{E}}_1 +
 \hat{j} \hat{\bm{E}}_2 + \hat{k} \hat{\bm{E}}_3)\delta x
\end{eqnarray}
with arbitrary integer numbers $\hat{i},\hat{j},\hat{k}$, while
a position vector in the rectangular coordinate system is defined as
\begin{eqnarray}
\bm{r}_{i,j,k} = (i \bm{e}_x + j \bm{e}_x + k \bm{e}_x)\delta x
\end{eqnarray}
with integer numbers $i,j,k$, where $\delta x$ represents the lattice spacing.
In general, the two position vectors 
$\hat{\bm{r}}_{\hat{i},\hat{j},\hat{k}}$ and $\bm{r}_{i,j,k}$
are not compatible with each other. 
To transform from $\hat{\bm{r}}_{\hat{i},\hat{j},\hat{k}}$ to
$\bm{r}_{i,j,k}$ using Eq. (\ref{transformationRtO}), $i$ must equal 
$\hat{i}-\gamma \hat{j}$. However, $\gamma \hat{j}$ is
not always an integer since $\gamma$ is defined between $0$ and $1$. 
We thus perform interpolation of the variables to overcome this problem.

\begin{figure}[t]
  \begin{center}
   \includegraphics[width=1.0\hsize]{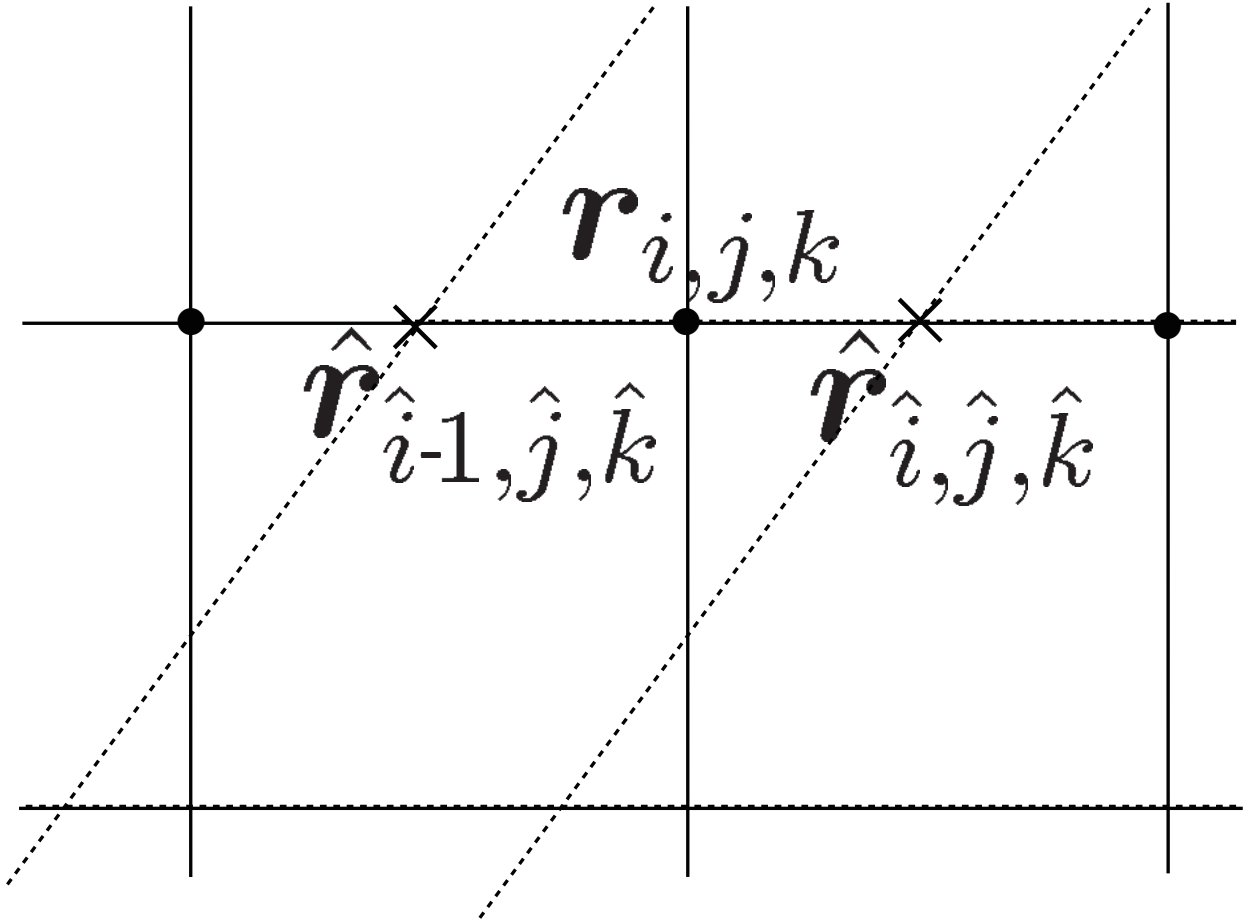}
   \caption{
A schematic illustration of the lattice discordance between the oblique
   and rectangular coordinates. $\bm{r}_{i,j,k}$ is the location vector in a
   rectangular coordinate system. $\hat{\bm{r}}_{\hat{i}-1,\hat{j},\hat{k}}$ and $\hat{\bm{r}}_{\hat{i},\hat{j},\hat{k}}$
   are the location vectors in the oblique coordinate system.
   \label{latttice_discordance}
   }
  \end{center}
 \end{figure}

Fig. \ref{latttice_discordance} shows a lattice discordance between the
rectangular and oblique coordinate systems. $\bm{r}_{i,j,k}$ is the
location vector in rectangular coordinates, and
$\hat{\bm{r}}_{\hat{i}-1,\hat{j},\hat{k}}$ and $\hat{\bm{r}}_{\hat{i},\hat{j},\hat{k}}$  are the
location vectors in oblique coordinates. Using liner interpolation, we
estimate the velocity field in rectangular coordinates
$\bm{u}(\bm{r}_{i,j,k})$ from the velocity field in oblique coordinates
$\bm{u}(\hat{\bm{r}}_{\hat{i}-1,\hat{j},\hat{k}})$ and
$\bm{u}(\hat{\bm{r}}_{\hat{i},\hat{j},\hat{k}})$. From the liner
interpolation, $\bm{u}(\bm{r}_{i,j,k})$ along the straight line is given by
the equation
\begin{eqnarray}
\bm{u}(\bm{r}_{i,j,k}) =
 \frac{|\hat{\bm{r}}_{\hat{i},\hat{j},\hat{k}} - \bm{r}_{i,j,k}|}
 {|\hat{\bm{r}}_{\hat{i},\hat{j},\hat{k}} - \hat{\bm{r}}_{\hat{i}-1,\hat{j},\hat{k}}|}\bm{u}(\hat{\bm{r}}_{\hat{i}-1,\hat{j},\hat{k}}) +
 \frac{|\hat{\bm{r}}_{\hat{i}-1,\hat{j},\hat{k}} - \bm{r}_{i,j,k}|}
 {|\hat{\bm{r}}_{\hat{i},\hat{j},\hat{k}} - \hat{\bm{r}}_{\hat{i}-1,\hat{j},\hat{k}}|}\bm{u}(\hat{\bm{r}}_{\hat{i},\hat{j},\hat{k}}).
\end{eqnarray}

\begin{figure}[t]
  \begin{center}
   \includegraphics[width=1.0\hsize]{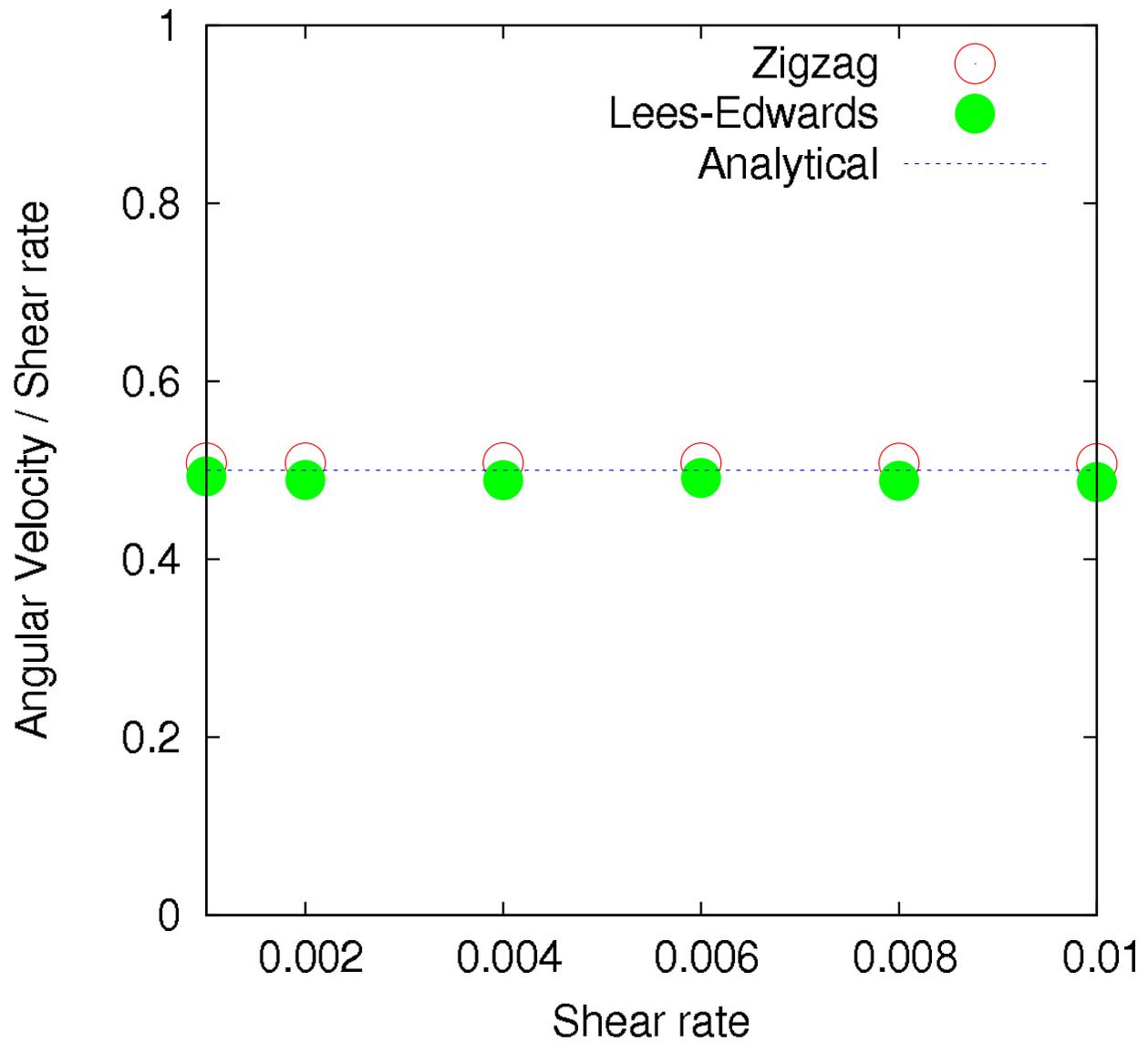}
   \caption{
The behavior of the ratio of angular velocity $\omega$ to the shear rate
   $\dot{\gamma}$ as a function of $\dot{\gamma}$. Open circles indicate
   the results of the previous method. Closed circles indicate the
   results of the present method. The solid line corresponds to the
   analytical solution.
   \label{compare_frq}
   }
  \end{center}
 \end{figure}

\begin{figure}[t]
  \begin{center}
   \includegraphics[width=1.0\hsize]{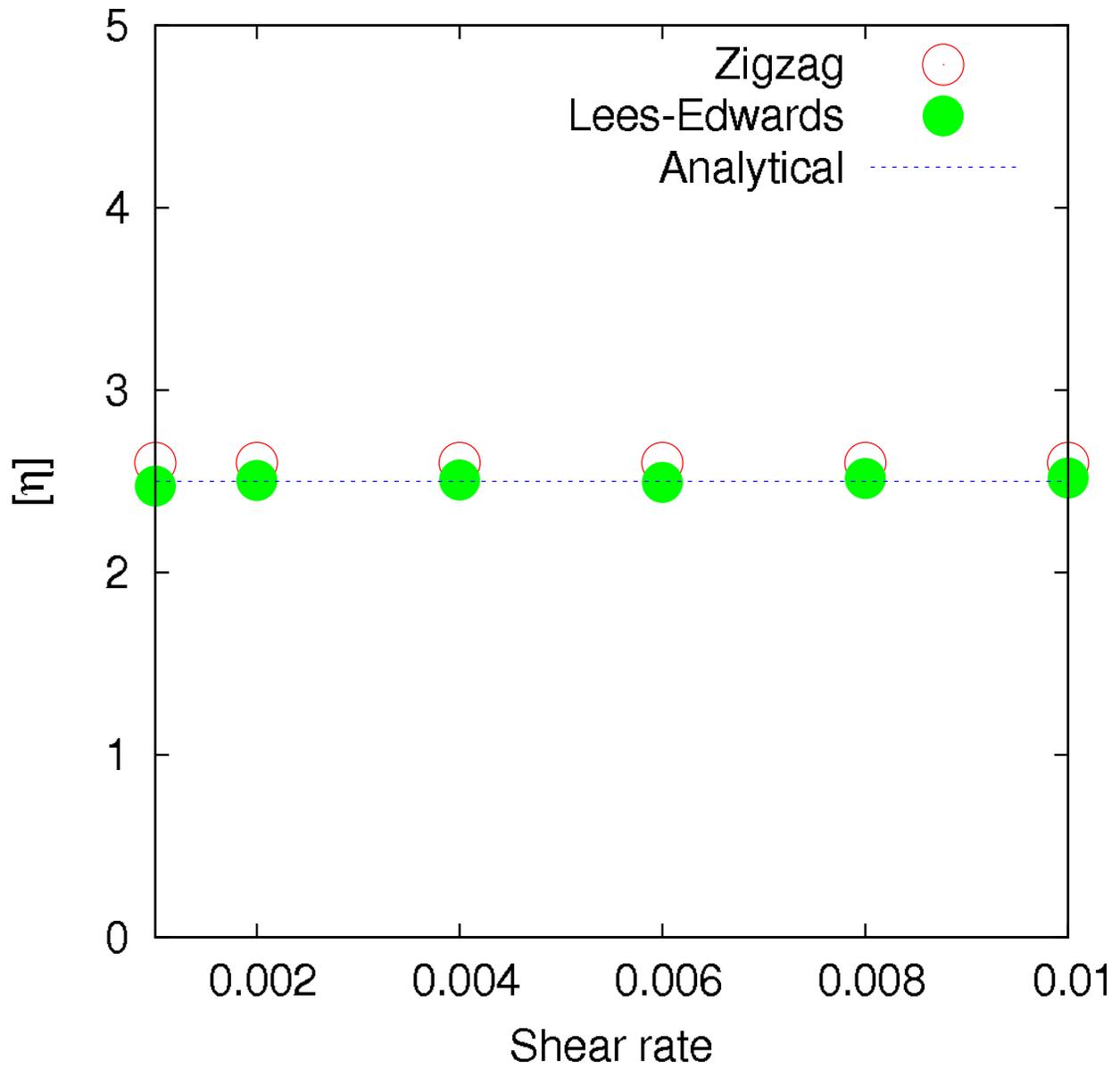}
   \caption{
The behavior of the intrinsic viscosity as a function of shear rate
   $\dot{\gamma}$. Open circles indicate the results of the previous
   method. Closed circles indicate the results of the present
   method. The solid line corresponds to Einstein's viscosity law.
   \label{compare_iv}
   }
  \end{center}
 \end{figure}

When using liner interpolation, artificial diffusion may arise. 
To check the reliability of the present method, we calculate
the angular velocity $\omega$ and intrinsic viscosity 
$[\eta] =\lim_{\Phi \to 0}(\eta - \eta_{\rm f})/\Phi$ 
for a dilute dispersion of
spherical particle for which analytical solutions are available.
As already shown in Figs. \ref{compare_frq} and \ref{compare_iv},
the present simulation data agree very well with analytical solutions
indicating that the effects of artificial numerical diffusion are not
serious.

%
%

\end{document}